\documentclass{ptephy}

\usepackage{braket}

\newcommand{\C}[1]{{\mathrm{C}_{#1}}}

\begin{document}

\title{Adiabatic internuclear potentials obtained by energy variation with the internuclear-distance constraint}

\author{
 \name{Yasutaka \textsc{Taniguchi} 谷口 億宇}{1,\ast},
 \name{Yoshiko \textsc{Kanada-En'yo} 延与 佳子}{2},
 and \name{Tadahiro \textsc{Suhara} 須原 唯広}{3}}

\address{
\affil{1}{Center for Computational Sciences, University of Tsukuba, Tsukuba, Ibaraki 305-8577, Japan}
\affil{2}{Department of Physics, Kyoto University, Kyoto, Kyoto 606-8502, Japan}
\affil{3}{Institute of Physics, University of Tsukuba, Tsukuba, Ibaraki 305-8577, Japan}
\email{yasutaka@ccs.tsukuba.ac.jp}
}

\begin{abstract}%
We propose a method to obtain adiabatic internuclear potentials via energy variation with the intercluster-distance constraint.
The adiabatic $^{16}$O+$^{16,18}$O potentials obtained by the proposed method are applied to investigate the effects of valence neutrons in $^{16}$O+$^{18}$O sub-barrier fusions.
 Sub-barrier fusion cross sections of $^{16}$O+$^{18}$O are enhanced more compared to those of $^{16}$O+$^{16}$O because of alignment of valence neutrons in $^{18}$O.
\end{abstract}

\subjectindex{D23, D26}

\maketitle

\section{Introduction}

Recent theoretical and experimental studies have revealed various exotic structures 
such as neutron halo and skin structures, where valence neutrons play important roles.
Nuclear reactions at low incident energy are efficient tools to investigate the effects of valence neutrons
in nuclear dynamics.
For instance, during tunneling through the Coulomb barrier, in sub-barrier nuclear fusion reactions, 
valence neutrons probably affect excitation of nuclei by influencing the alignment, polarization, and vibration. 
To investigate low-energy nuclear reactions theoretically, internuclear potentials that work well in low-incident energy are required.

There are two types of internuclear potentials for nuclear reactions; adiabatic and sudden potentials.
To obtain adiabatic potentials in mean-field approaches, two methods have been proposed: applications of the time-dependent Hartree--Fock (HF) method with a density constraint\cite{PhysRevC.74.021601} and the HF method with an internuclear-distance constraint in a symmetric form\cite{Zint1975269,PhysRevC.14.1488}.
In the former method, energy is minimized using the constraint on the density distributions obtained by the time-dependent HF calculations.
Since the density is determined by time-dependent HF at incident energies greater than the Coulomb barrier, the
distortion effects of colliding nuclei associated with density changes 
at low incident energy (i.e., below the Coulomb barrier) can be insufficient. 
In the latter method, energy variation with a symmetric constraint on the internuclear distances has been applied to symmetric systems but not to asymmetric systems.
Double-folding potential, which is a type of sudden potential, has also been used to study nuclear reactions; fusion cross sections near the Coulomb barrier are described by 
channel-coupling (CC) calculations with double-folding potentials\cite{PhysRevC.69.024610}.
Recently, a repulsive core potential has been suggested 
to account for deep sub-barrier fusion cross sections\cite{PhysRevC.75.034606}.
However, the fundamental origin of the phenomenological repulsive effect has not been clarified yet.

In sub-barrier nuclear fusion, adiabatic potentials are expected to work, which contain excitation effects of colliding nuclei.
Adiabatic potentials treating excitation effects of colliding nuclei contains CC effects approximately\cite{PhysRevLett.31.766,PhysRevC.34.894,RevModPhys.70.77,doi:10.1146/annurev.nucl.48.1.401}.
To investigate the excitation effects of valence neutrons
on fusion cross sections, comparison of $^{16}$O+$^{16,18}$O sub-barrier fusions are feasible because $^{18}$O possesses two neutrons more than $^{16}$O.
The sub-barrier fusion cross sections of $^{16}$O+$^{18}$O are enhanced more compared with those 
of $^{16}$O+$^{16}$O although the enhancement is smaller than that in heavier systems such as Ca isotopes\cite{PhysRevC.31.1980}.
The CC model shows that most of the enhancement is attributable to the excitation of the low $2^+$ state in $^{18}$O\cite{PhysRevC.32.1432}.
In potential model interpretation, it means that the excess neutrons result in an effectively thinner or lower Coulomb barrier for $^{16}$O+$^{18}$O than that for $^{16}$O+$^{16}$O.
The CC model is a popular to analyze fusion reactions around the Coulomb barrier and gives reasonable results\cite{RevModPhys.70.77}.
However, most of internuclear potentials used in the CC model are
phenomenological ones, and they are not based on microscopic frameworks
where the antisymmetrization effects between colliding nuclei are taken
into account.

This paper aims to propose a method to obtain adiabatic internuclear potentials in a full-microscopic framework.
In the process, the energy is minimized while constraining the internuclear distances via the deformed-basis antisymmetrized molecular dynamics  (AMD) model, which can be easily applied to both asymmetric and symmetric systems.
The adiabatic $^{16}$O+$^{18}$O potentials obtained by the proposed method show that the excitation of valence neutrons in $^{18}$O reduces the thickness of the Coulomb barriers and enhances the sub-barrier fusion cross sections of $^{16}$O+$^{18}$O.
The excitation effects are analyzed via comparison of cross sections obtained by adiabatic and sudden potentials.
In general, the enhancement effects of sub-barrier fusion cross sections are larger in heavier systems but we choose the light systems $^{16}$O+$^{16,18}$O, due to the numerical cost for obtaining the internuclear potentials.

In Sec.~\ref{framework}, we explain the framework to obtain internuclear potentials and fusion cross sections.
In Sec.~\ref{results}, we present the internuclear potentials and fusion cross sections.
In Sec.~\ref{discussions}, we discuss the role of valence neutrons in $^{18}$O to enhance sub-barrier fusion cross sections.
Finally, conclusions are given in Sec.~\ref{conclusions}.

\section{Framework}
\label{framework}

We propose a method to obtain adiabatic potentials via energy variation with the intercluster-distance constraint in the AMD framework\cite{PTP.112.475}.
In the present study, we use the deformed-basis AMD framework\cite{PhysRevC.69.044319}.
A form of the deformed-basis AMD wave function $| \mathrm{\Phi} \rangle$, Slater determinant of Gaussian wave packets, is described as
\begin{equation}
 | \mathrm{\Phi} \rangle = \hat{\mathcal{A}} | \varphi_1, \varphi_2, ..., \varphi_A \rangle,
\end{equation}
\begin{equation}
 | \varphi_i \rangle = | \phi_i \rangle \otimes | \chi_i \rangle,
\end{equation}
\begin{equation} 
\langle \mathbf{r} | \phi_i \rangle = \pi^{-\frac{3}{4}} (\det \mathsf{K})^\frac{1}{2} \exp\left[ - \frac{1}{2} ( \mathsf{K} \mathbf{r} - \mathbf{Z}_i)^2\right],
\end{equation}
where $\hat{\mathcal{A}}$ is the antisymmetrization operator, and $| \phi_i \rangle$ and $| \chi_i \rangle$ are spatial and spin-isospin parts, respectively.
$\mathsf{K}$ is a real $3 \times 3$ matrix that denotes the width of Gaussian wave packets, which is common to all nucleons, and $\mathbf{Z}_i$ is a complex vector that denotes a centroid of a Gaussian wave packet in phase space.
The wave function is set as the expectation values of the position of center of mass and the total momentum are zero.
A wave function $| \mathrm{\Phi}_{\C{1}-\C{2}} \rangle$, having a dinuclear structure comprising nuclei $\C{1}$ and $\C{2}$, is defined as
\begin{equation}
 | \mathrm{\Phi}_{\C{1}-\C{2}} \rangle = \hat{\mathcal{A}} \left( | \mathrm{\Phi}_\C{1} \rangle \otimes | \mathrm{\Phi}_\C{2} \rangle \right) ,
\end{equation}
where $| \mathrm{\Phi}_\C{i} \rangle$ is a direct product, which is not antisymmetrized, of single-particle wave functions with proton and neutron numbers corresponding to the nucleus $\C{i}$.

The internuclear distance $R$ is defined by the density distribution of a total system.
Suppose the centers of mass of nuclei $\C{1}$ and $\C{2}$ are located on the $z$-axis with $z < 0$ and $z > 0$, respectively.
Boundary planes of nuclei $\C{1}$ and $\C{2}$ for protons and neutrons are denoted by $z = z_p$ and $z = z_n$, respectively, and are defined as
\begin{equation}
 \int_{-\infty}^{z_p}\!\!dz \int\!\!\!\!\int dxdy\ \rho_p(\mathbf{r}) = Z_1,
\end{equation}
\begin{equation}
 \int_{-\infty}^{z_n}\!\!dz \int\!\!\!\!\int dxdy\ \rho_n(\mathbf{r}) = N_1.
\end{equation}
Here $\rho_p(\mathbf{r})$ and $\rho_n(\mathbf{r})$ denote proton and neutron densities, and $Z_1$ and $N_1$ denote proton and neutron numbers of nucleus C$_1$, respectively.
The internuclear distance $R$ is defined by the positions $\mathbf{R}_1$ and $\mathbf{R}_2$ of the centers of mass of nuclei $\C{1}$ and $\C{2}$, respectively, as
\begin{equation}
 R = \left|\mathbf{R}_2 - \mathbf{R}_1\right|,
\end{equation}
\begin{equation}
 \mathbf{R}_i = \frac{Z_i \mathbf{R}^{(p)}_i + N_i \mathbf{R}^{(n)}_i}{A_i},
\end{equation}
\begin{equation}
 \mathbf{R}^{(p,n)}_1 = \int\!\!\!\!\int dxdy\int_{-\infty}^{z_{p,n}}\!\!dz\ \mathbf{r}~\rho_{p,n} (\mathbf{r}),\label{Rpn1}
\end{equation}
\begin{equation}
 \mathbf{R}^{(p,n)}_2 = \int\!\!\!\!\int dxdy\int^{\infty}_{z_{p,n}}\!\!dz\ \mathbf{r}~\rho_{p,n} (\mathbf{r}),\label{Rpn2}
\end{equation}
where $Z_i$, $N_i$, and $A_i$ denote the proton, neutron, and mass number of a nucleus C$_i$, respectively.

 To obtain the adiabatic potentials $V_{\mathrm{ad}}$, we optimize the dinuclear wave function while constraining the $\C{1}$--$\C{2}$ distance using the $d$-constraint AMD method\cite{PTP.112.475}. That is, the energy is minimized while constraining the distance parameter $d$ between the centers of mass of the wave packets of nuclei $\C{1}$ and $\C{2}$ according to 
\begin{eqnarray}
 &\delta\left[\langle \mathrm{\Phi}_{\C{1}-\C{2}}^{\mathrm{(opt)}}; d| \hat{H}' |\mathrm{\Phi}_{\C{1}-\C{2}}^{\mathrm{(opt)}}; d \rangle + V_{\mathrm{cnst}} (d)\right] = 0,&\label{variation}\\
 &\hat{H}' = \hat{T} + \hat{V}_{\mathrm{N}} + \hat{V}_{\mathrm{C}} - 2 \hat{T}_{\mathrm{G}},&\label{internuclear_potential}
\end{eqnarray}
where $\hat{T}$, $\hat{V}_{\mathrm{N}}$, $\hat{V}_{\mathrm{C}}$ and $\hat{T}_{\mathrm{G}}$ are the kinetic energy, effective nuclear interaction, Coulomb interaction, and kinetic energy of the center-of-mass motion of the total system, respectively. 
$V_{\mathrm{cnst}} (d)$ denotes a parabolic constraint potential for the internuclear distance $d$ defined by using the set of single-particle wave functions $| \mathrm{\Phi}_{\C{i}} \rangle$:
\begin{equation}
V_{\mathrm{cnst}} (d) = v_{\mathrm{cnst}} 
  \left[~
   \left| 
    \langle \mathrm{\Phi}_\C{1} | \hat{\mathbf{R}}_{\mathrm{G}1} \hat{\mathcal{A}} | \mathrm{\Phi}_\C{1} \rangle
    -
    \langle \mathrm{\Phi}_\C{2} | \hat{\mathbf{R}}_{\mathrm{G}2} \hat{\mathcal{A}} | \mathrm{\Phi}_\C{2} \rangle
   \right|
   -
   d~
  \right]^2,
\end{equation}
\begin{equation}
  \hat{\mathbf{R}}_{\mathrm{G}i} \equiv \frac{1}{A_i} \sum_{j = 1}^{A_i} \hat{\mathbf{r}}_{j},
\end{equation}
where $v_{\mathrm{cnst}}$ denotes a sufficiently large number.
Details of the constraint potential are reported in Ref.\cite{PTP.112.475}.
In $d \gtrsim 6$ fm, the distance $R$ defined by the density distribution agrees with $d$ defined by the centers of mass of subsystems for the $^{16}$O+$^{16}$O system\cite{PTP.128.349}.
Adiabatic potentials reflect structural changes with respect to internuclear distances.
By using the optimized wave function $| \mathrm{\Phi}_{\C{1}-\C{2}}^{\mathrm{(opt)}}; d \rangle$, an adiabatic potential $V_{\mathrm{ad}} (R)$ is defined as
\begin{equation}
 V_{\mathrm{ad}} (R(d)) = \langle \mathrm{\Phi}_{\C{1}-\C{2}}^{\mathrm{(opt)}}; d | \hat{H}' | \mathrm{\Phi}_{\C{1}-\C{2}}^{\mathrm{(opt)}}; d \rangle
  - (E_{\C{1}\mathrm{gs}} + E_{\C{2}\mathrm{gs}}),
\end{equation}
where $E_{\C{1}\mathrm{gs}} + E_{\C{2}\mathrm{gs}}$ denotes a summation of ground-state energies of nuclei $\C{1}$ and $\C{2}$ obtained by varying the energy for isolated systems $\C{1}$ and $\C{2}$ in case of common width matrices $\mathsf{K}$ for the wave functions of $\C{1}$ and $\C{2}$.
Wave functions obtained by the energy variation for a summation of energies of nuclei $\C{1}$ and $\C{2}$ are denoted as $| \mathrm{\Phi}_\C{i}^{\mathrm{(gs)}} \rangle\ (i = 1, 2)$.

To analyze excitation effects of colliding nuclei, we also define the sudden potentials $V_{\mathrm{sud}} (R)$.
We use dinuclear wave functions $|\mathrm{\Phi}_{\C{1}-\C{2}}^{\mathrm{(gs)}}; R \rangle$
defined by ground-state wave functions of nuclei $\C{1}$ and $\C{2}$. 
The wave functions $|\mathrm{\Phi}_{\C{1}-\C{2}}^{\mathrm{(gs)}}; R \rangle$ are defined by shifting the ground-state wave functions $| \mathrm{\Phi}_\C{i}^{\mathrm{(gs)}} \rangle\ (i = 1, 2)$ to a certain position such that the internuclear distance is equal to $R$, and the total wave function is antisymmetrized.
Thus, the structures of nuclei $\C{1}$ and $\C{2}$ are frozen, except for the effects of antisymmetrization between nuclei $\C{1}$ and $\C{2}$.
Next, we define the sudden potential as, 
\begin{equation}
 V_{\mathrm{sud}} (R) = \langle \mathrm{\Phi}_{\C{1}-\C{2}}^{\mathrm{(gs)}}; R | \hat{H}' | \mathrm{\Phi}_{\C{1}-\C{2}}^{\mathrm{(gs)}}; R \rangle
  - (E_{\C{1}\mathrm{gs}} + E_{\C{2}\mathrm{gs}}).
\end{equation}
For the $^{16}$O+$^{18}$O system, since the ground-state wave function of $^{18}$O is deformed, the orientation $\Omega$ of $^{18}$O is averaged to obtain the
sudden potential:
\begin{equation}
 V_{\mathrm{sud}}(R) = \frac{1}{4 \pi} \int V'_{\C{1}-\C{2}}(R, \Omega) d\Omega,
\end{equation}
\begin{equation}
 V'_{\C{1}-\C{2}}(R, \Omega)  = \langle \mathrm{\Phi}^{\mathrm{(gs)}}_{\C{1}-\C{2}}; R, \Omega | \hat{H}' | \mathrm{\Phi}^{\mathrm{(gs)}}_{\C{1}-\C{2}}; R, \Omega \rangle  - (E_{\C{1}\mathrm{gs}} + E_{\C{2}\mathrm{gs}}).
\end{equation}
For practical purposes, $\Omega$ integration is achieved by averaging the direction of 
the shift in the position of nuclei $\C{1}$ and $\C{2}$.

The data points for the potentials are calculated at intervals of approximately 0.5 fm and are interpolated by spline curves to obtain potentials as functions of $R$.

To obtain the present internuclear potentials, we adopt a distinct treatment of subtracting $2 \hat{T}_{\mathrm{G}}$ instead of $\hat{T}_{\mathrm{G}}$ from the Hamiltonian [Eq.~(\ref{internuclear_potential})].
The $2\hat{T}_{\mathrm{G}}$ is subtracted to eliminate the kinetic energy of the internuclear motion $T_\mathrm{rel}$ and that of the center-of-mass motion $T_\mathrm{G}$ of the total system. 
The expectation value $T_i$ of the kinetic energy of the center-of-mass motion of nucleus $\C{i}$ is separated into the classical part $T^{\mathrm{(cl)}}_i = \frac{\mathbf{P}^2_i}{2 A_i m}$ and the other part $T_{\mathrm{G}i}$ as
\begin{equation}
 T_i = T^{\mathrm{(cl)}}_i + T_{\mathrm{G}i},
\end{equation}
here $\mathbf{P}_i$ is the expectation value of the total momentum of the nucleus $\C{i}$.
When two nuclei $\C{1}$ and $\C{2}$ are well separated, $T_{\mathrm{G}1}$ and $T_{\mathrm{G}2}$ are functions of $\mathsf{K}$ as
\begin{equation}
 T_{\mathrm{G}1} = T_{\mathrm{G}2} = \frac{\hbar^2}{2m} \mathrm{tr} \left( \frac{1}{2} {}^t\mathsf{K}\mathsf{K} \right),
\end{equation}
which equals to $T_\mathrm{G}$.
For the adiabatic condition $\mathbf{P}_1 = \mathbf{P}_2 = \mathbf{0}$, wave functions of subsystems are set as $T_i^\mathrm{(cl)} = 0$ in obtaining sudden potentials.
In obtaining adiabatic potentials, A $\left(T_1^\mathrm{(cl)} + T_2^\mathrm{(cl)}\right)$ term is used as a constraint potential for the adiabatic condition in the energy variation, and resultant value of $T_i^\mathrm{(cl)}$ is small and negligible.
It gives
\begin{equation}
 T_1 = T_2 = T_\mathrm{G}
\end{equation}
in both cases of adiabatic and sudden potentials.
Using the equations, the $2T_\mathrm{G}$ term is written as a summation of the $T_\mathrm{G}$ and $T_\mathrm{rel}$ as
\begin{equation}
2T_\mathrm{G} = T_1 + T_2 = T_\mathrm{G} + T_\mathrm{rel},
\end{equation}
and the expectation value of $\hat{H}'$ is written as
\begin{equation}
 \braket{\hat{H}'} = \braket{\hat{T}} + \braket{\hat{V}_{\mathrm{N}}} + \braket{\hat{V}_{\mathrm{C}}} - T_\mathrm{G} - T_\mathrm{rel}.
\end{equation}
Although the $T_{\mathrm{G}1}$ and $T_{\mathrm{G}2}$ values deviate from $T_\mathrm{G}$ in the overlap region because of antisymmetrization, this effect is small in the barrier region because overlap between nuclei $\C{1}$ and $\C{2}$ is small in the region.
Hence, $2\hat{T}_{\mathrm{G}}$ is subtracted in the present calculations.
By these definitions, the internuclear potentials indicate Coulomb potentials in large $R$ region.

The Modified Volkov No.1 case 1\cite{PTP.64.1608} and Gogny D1S (D1S) interactions are used as effective nuclear interactions $\hat{V}_{\mathrm{N}}$.
In the Modified Volkov No.1 interaction, a three-body contact term is replaced with a density-dependent two-body term, and a spin-orbit term of the D1S interaction is added to adjust the threshold energy of $^{32}$S to that of $^{16}$O+$^{16}$O (MV1$'$).

To obtain fusion cross sections, we use the potential model code \textsc{potfus3} provided by Hagino {\it et al}.\cite{Hagino1999143}.
The \textsc{potfus3} directly integrates second order differential equations using the modified Numerov method to solve the Schr\"odinger equation.
 Inside the Coulomb barrier, the incoming wave boundary conditions that there are only incoming waves at $r = r_\mathrm{min}$ is employed.
The $r_\mathrm{min}$ is set to 6 fm in the present calculations.

\section{Results}
\label{results}

\begin{figure}[tbp]
 \begin{center}
\begin{tabular}{cc}
 \includegraphics[width=0.475\textwidth]{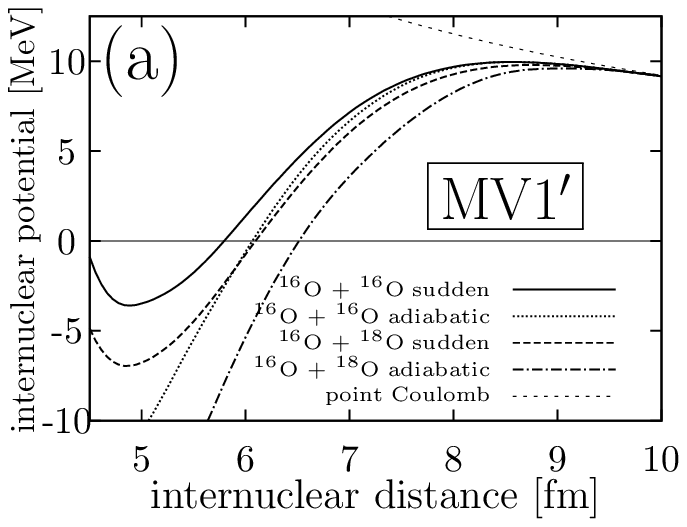}&
 \includegraphics[width=0.475\textwidth]{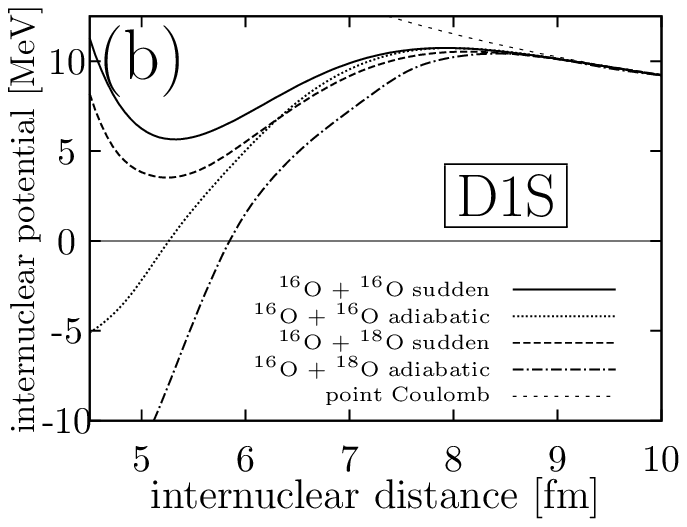}
\end{tabular}  
\caption{
  Adiabatic and sudden $^{16}$O+$^{16,18}$O potentials calculated as functions of the internuclear distance for (a) the MV1$'$ and (b) D1S  interactions.
  Solid and dotted lines represent sudden and adiabatic $^{16}$O+$^{16}$O potentials, respectively; short-dashed and dot-dashed lines represent sudden and adiabatic $^{16}$O+$^{18}$O potentials, respectively; and the long-dashed line represents the point Coulomb potential.
  }
  \label{potential}
 \end{center}
\end{figure}

Figure~\ref{potential} shows the adiabatic and sudden $^{16}$O+$^{16,18}$O potentials calculated for the MV1$'$ and D1S interactions.
The $(T^{\mathrm{(cl)}}_1 + T^\mathrm{(cl)}_2)$ value is less than $5 \times 10^{-2}$ MeV for $R \geq 4.5$ fm region.
The two interactions result in qualitatively similar internuclear potentials.
Each sudden potential has a structural repulsive core\cite{PTP.34.191} in the $R \lesssim 5$ fm region because of the Pauli blocking. In the $R \gtrsim 6$ fm region, the adiabatic and sudden $^{16}$O+$^{16}$O potentials are similar to each other.
Both potentials show barrier tops at almost the same internuclear distances, and the shape of the potential curves is also similar within the barrier top. 
The sudden $^{16}$O+$^{18}$O potential is similar to the adiabatic and sudden $^{16}$O+$^{16}$O potentials.
However, the adiabatic $^{16}$O+$^{18}$O potential is lower than other potentials.
Due to the lower nuclear potential, the barrier tops of the adiabatic $^{16}$O+$^{18}$O potentials occur at larger internuclear distances. The difference between the sudden and adiabatic potentials
indicates that the effect of excitation is large in the $^{16}$O+$^{18}$O system, whereas 
it is small in the $^{16}$O+$^{16}$O system.
In both potentials, the MV1$'$ interaction gives lower potentials as compared with those given by the D1S interaction.

\begin{figure}[tbp]
 \begin{center}
  \begin{tabular}{cc}
   \includegraphics[width=0.25\textwidth]{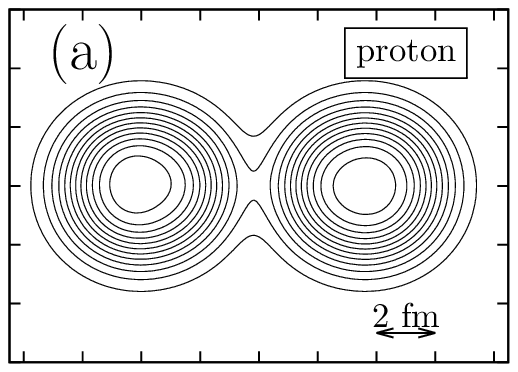} &
   \includegraphics[width=0.25\textwidth]{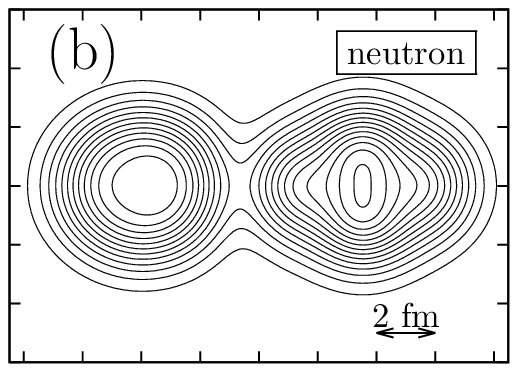}\\
   &\includegraphics[width=0.25\textwidth]{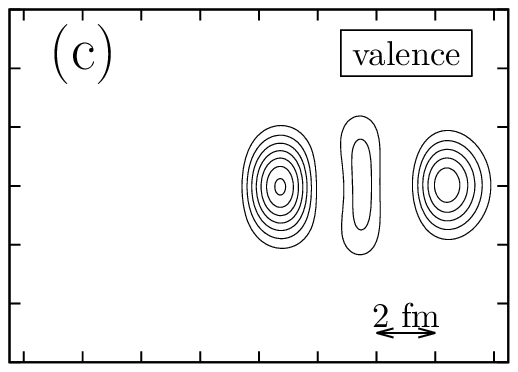}
  \end{tabular}  
  \caption{
  Density distribution of (a) protons, (b) neutrons, and (c) valence neutrons of $^{16}$O+$^{18}$O at $R = 7.5$ fm obtained by varying the energy for the MV1$'$ interaction.
  }
  \label{density}
 \end{center}
\end{figure}

Figure~\ref{density} shows the density distribution of the proton and neutron parts in the 
$^{16}$O+$^{18}$O wave functions at $R = 7.5$ fm. 
This distribution is obtained by varying the energy with a constraint on the internuclear distance for the MV1$'$ interaction.
The density distribution of the two valence neutrons, which is defined by subtracting the proton density from the neutron density assuming density distributions of protons and neutrons are similar for the $^{16}$O core,
is also shown in the figure. 
The valence neutron density in $^{18}$O has three peaks, which is a result similar to that of the intrinsic 
wave functions of the $^{18}$O ground state although the $^{18}$O ground state has a spherical density distribution in the laboratory frame.

\begin{figure}[tbp]
 \begin{center}
\begin{tabular}{cc}
 \includegraphics[width=0.475\textwidth]{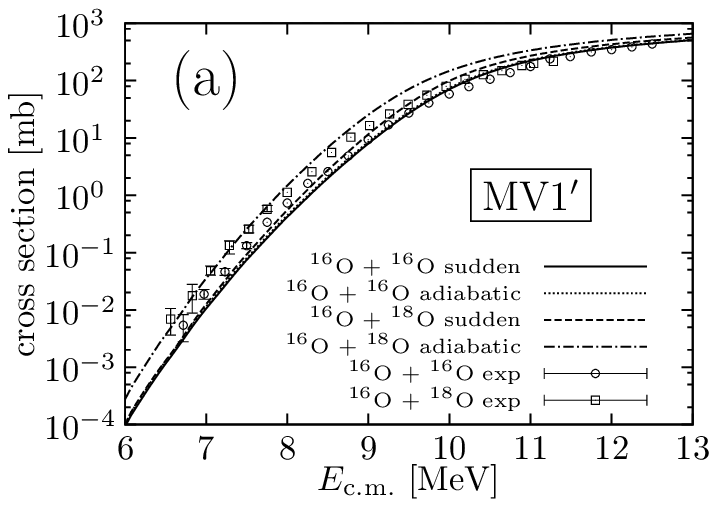}&
 \includegraphics[width=0.475\textwidth]{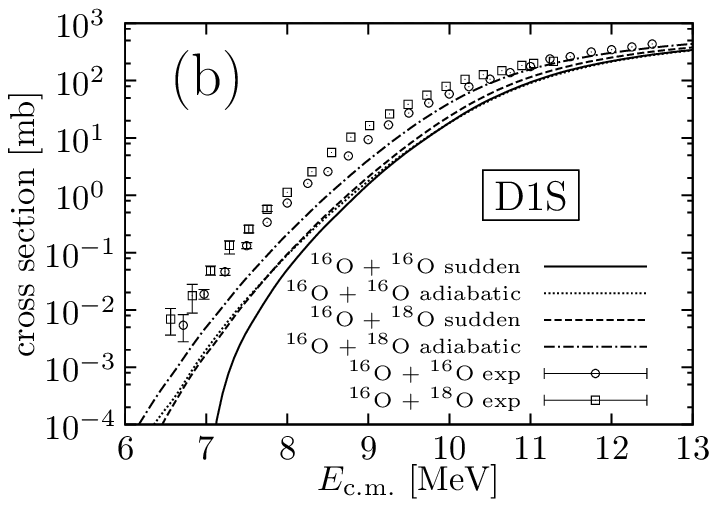}
\end{tabular}  
\caption{
  Fusion cross section for $^{16}$O+$^{16}$O  and $^{16}$O+$^{18}$O  as a function of the incident energy in the center-of-mass frame obtained by adiabatic and sudden potentials for (a) the MV1$'$ and (b) D1S interactions.
  Solid and dotted lines are obtained by sudden and adiabatic $^{16}$O+$^{16}$O potentials, respectively, and short-dashed and dot-dashed lines are obtained by sudden and adiabatic $^{16}$O+$^{18}$O potentials, respectively.
 The circles and squares denote experimental values for $^{16}$O+$^{16}$O and $^{16}$O+$^{18}$O fusion cross sections, respectively, which are taken from Ref.\cite{PhysRevC.31.1980}.
  }
  \label{cross_section}
 \end{center}
\end{figure}

We calculated the $^{16}$O+$^{16,18}$O fusion cross sections with the 
adiabatic and sudden potentials using the potential model code \textsc{potfus3}\cite{Hagino1999143}.
Figure~\ref{cross_section} shows the $^{16}$O+$^{16,18}$O fusion cross sections
as functions of incident energy in the center-of-mass frame. 
The energy dependence of the $^{16}$O+$^{16,18}$O sub-barrier cross sections 
for $E_{\mathrm{c.m.}} \lesssim 9$ MeV, 
obtained with sudden and adiabatic potentials calculated using the MV1$'$ and D1S interactions, 
has qualitatively similar slopes,
except for the $^{16}$O+$^{16}$O fusion cross section calculated using the D1S interaction, which decreases rapidly
for $E_{\mathrm{c.m.}} \lesssim 8$ MeV.
The $^{16}$O+$^{16}$O sub-barrier fusion cross sections $\sigma_{\mathrm{ad}}(^{16}\mathrm{O})$ and $\sigma_{\mathrm{sud}}(^{16}\mathrm{O})$ obtained with adiabatic and sudden potentials, 
respectively, are similar to each other.
The $^{16}$O+$^{18}$O sub-barrier fusion cross section $\sigma_{\mathrm{ad}}(^{18}\mathrm{O})$ obtained by adiabatic potentials is a few times larger than the 
$^{16}$O+$^{16}$O sub-barrier fusion cross sections $\sigma_{\mathrm{ad}}(^{16}\mathrm{O})$ and $\sigma_{\mathrm{sud}}(^{16}\mathrm{O})$ because of the lower internuclear potentials.
However, the cross section $\sigma_{\mathrm{sud}}(^{18}\mathrm{O})$, obtained by sudden potentials,
is smaller than $\sigma_{\mathrm{ad}}(^{18}\mathrm{O})$ but  
similar to $\sigma_{\mathrm{ad}}(^{16}\mathrm{O})$ and $\sigma_{\mathrm{sud}}(^{16}\mathrm{O})$. 
For the MV1$'$ interaction, the cross sections $\sigma_{\mathrm{ad}}(^{16}\mathrm{O})$ and $\sigma_{\mathrm{ad}}(^{18}\mathrm{O})$ agree with the experimental data qualitatively, whereas for the D1S interactions, the estimated sub-barrier fusion cross sections are less than the experimental data.

\section{Discussions}
\label{discussions}

In this section, we discuss the contribution of valence neutrons in enhancing the sub-barrier fusion cross sections 
by analyzing the results obtained for the MV1$'$ interaction, which accounts for the measured $^{16}$O+$^{16,18}$O sub-barrier fusion cross sections qualitatively.
As mentioned previously, the experimental enhancement of $^{16}$O+$^{18}$O sub-barrier fusion cross sections compared with $^{16}$O+$^{16}$O fusion cross sections
is reproduced by the adiabatic potentials having lower internuclear potential  than the sudden potentials.
As shown in Fig.~\ref{cross_section}, 
the $^{16}$O+$^{16,18}$O sub-barrier fusion cross sections obtained with adiabatic and sudden potentials are related as follows:
\begin{equation}
 \sigma_{\mathrm{sud}}(^{16}\mathrm{O}) \sim \sigma_{\mathrm{ad}}(^{16}\mathrm{O}) \sim \sigma_{\mathrm{sud}}(^{18}\mathrm{O}) < \sigma_{\mathrm{ad}}(^{18}\mathrm{O}).
\end{equation}
The relation $\sigma_{\mathrm{sud}}(^{16}\mathrm{O}) \sim \sigma_{\mathrm{ad}}(^{16}\mathrm{O})$
indicates that the effects of $^{16}$O excitation on sub-barrier fusion cross section are negligible.
However, for $^{16}$O+$^{18}$O, sub-barrier fusion cross sections obtained with adiabatic potentials, which is significantly larger than those obtained with sudden potentials, 
indicate 
that the excitation of $^{18}$O enhances the $^{16}$O+$^{18}$O fusion cross sections. The relation $\sigma_{\mathrm{sud}}(^{16}\mathrm{O}) \sim \sigma_{\mathrm{ad}}(^{16}\mathrm{O}) \sim \sigma_{\mathrm{sud}}(^{18}\mathrm{O})$ 
implies that when the $^{18}$O nucleus is frozen, the theory cannot account 
for the enhancement of the $^{16}$O+$^{18}$O sub-barrier fusion cross sections.
Since excitation of the $^{16}$O core is minor, the excitation of $^{18}$O is primarily due to the excitation of two valence neutrons around the $^{16}$O core.
In other words, the excitation of the two valence neutrons in $^{18}$O enhances the sub-barrier fusion cross sections, whereas increasing the number of neutrons without excitation does not have significant contribution. 

 To investigate the details of the excitation effects of valence neutrons in $^{18}$O, we discuss the alignment and dipole polarization. 
Since the $^{18}$O ground state exhibits an intrinsic deformation, 
one of the possible excitation effects is the alignment of deformed $^{18}$O. Dipole polarization can be another
excitation effect. 
The neutrons in $^{18}$O may distribute inward (toward $^{16}$O) relative to protons because of the isospin dependence of nuclear interactions and the Coulomb force, which results in the isovector dipole polarization of $^{18}$O.
Analysis of the alignment effects and dipole polarization reveals that the alignment effect
contributes significantly to the enhancement of the $^{16}$O+$^{18}$O sub-barrier fusion cross sections.

\begin{figure}[tbp]
 \begin{center}
  \includegraphics[width=.5\textwidth]{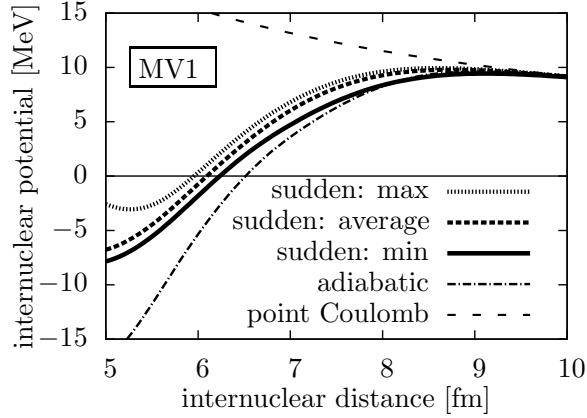}
  \caption{
  Maximum (dotted) and minimum (solid) $^{16}$O+$^{18}$O internuclear potentials for the orientation of $^{18}$O, in which $^{16}$O and $^{18}$O are frozen, as functions of internuclear distance for the MV1$'$ interaction.
  The sudden (short dashed) and adiabatic (dot-dashed) $^{16}$O+$^{18}$O potentials are also shown.
  The long-dashed line represents point Coulomb potential.
  }
  \label{alignment}
 \end{center}
\end{figure}

For calculating $^{16}$O+$^{18}$O sudden potentials, we assumed that nuclei $^{16}$O and $^{18}$O are frozen, and 
the orientation $\Omega$ of $^{18}$O is averaged.
 To study the effect of the alignment of the deformed $^{18}$O on $^{16}$O+$^{18}$O potentials, we analyzed them before averaging the $^{18}$O orientation $\Omega$. 
 At every internuclear distance, the
orientation $\Omega$ is optimized to obtain the minimum (maximum) $^{16}$O+$^{18}$O energy $V'_{^{16}\mathrm{O}-^{18}\mathrm{O}}(R, \Omega)$, from which the minimum (maximum) $^{16}$O+$^{18}$O potential is calculated with frozen $^{16}$O and $^{18}$O.
The results are shown in Fig.~\ref{alignment} as functions of the internuclear distance. 
The $^{16}$O+$^{18}$O sudden potential (after averaging $\Omega$) and the adiabatic potential are also shown for comparison.
In the $R \gtrsim 6$ fm region, the sudden potential is similar to the maximum internuclear potential.
In the $R \gtrsim 7$ fm region, the $^{16}$O+$^{18}$O adiabatic potential is similar to the minimum internuclear potential, in which $^{16}$O and $^{18}$O are frozen. 
In particular, the minimum values of $^{16}$O+$^{18}$O potential give almost the same barrier height and thickness as those of the 
$^{16}$O+$^{18}$O adiabatic potential, at least for fusion at $E_{\mathrm{c.m.}} \ge 7$ MeV. 
These results indicate that the alignment of $^{18}$O majorly describes the difference between the  sudden and adiabatic potentials.
 We conclude that the enhancement of the $^{16}$O+$^{18}$O sub-barrier fusion cross section is due to the alignment effect.

\begin{figure}[tbp]
 \begin{center}
  \includegraphics[width=0.5\textwidth]{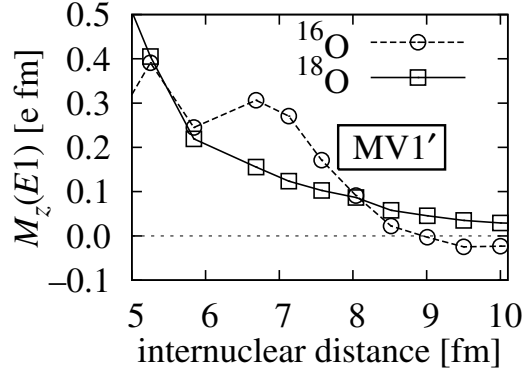}
  \caption{Dipole moments of $^{16}$O (dashed) and $^{18}$O (solid) in $^{16}$O+$^{18}$O structures obtained by varying the energy for the MV1$'$ interaction.}
  \label{dipole}
 \end{center}
\end{figure}

Dipole polarization can be another effect of excitation in $^{18}$O.
Here the isovector dipole moments $\mathbf{M}(E1)$ are defined as,
\begin{equation}
 \mathbf{M}(E1) = \sqrt{\frac{3}{4 \pi}} \frac{NZe}{A}(\mathbf{R}_p -\mathbf{R}_n),
\end{equation}
where $\mathbf{R}_p$ and $\mathbf{R}_n$ denote the positions of the centers of mass of protons and neutrons, respectively, and are  defined in Eqs.~(\ref{Rpn1}) and (\ref{Rpn2}).
In the $^{16}$O+$^{18}$O system, $^{18}$O valence neutrons are somewhat 
polarized because the existence of $^{16}$O results in finite isovector dipole moments for $^{18}$O.
Figure~\ref{dipole} shows $z$-components of the dipole moments $M_z(E1)$ of $^{16}$O and $^{18}$O in the $^{16}$O+$^{18}$O wave functions as functions of the internuclear distance $R$, where $z$-axis is a major axis of the total systems.
$|M_x(E1)|$ and $|M_y(E1)|$ are considerably smaller than $|M_z(E1)|$.
The $^{16}$O and $^{18}$O nuclei are located at $z < 0$ and $z > 0$, respectively.
 The $^{16}$O+$^{18}$O wave functions are obtained by the $d$-constraint 
AMD method using the MV1$'$ interaction [Eq.~(\ref{variation})]. 
In the large-distance region $R \gtrsim 9$ fm, $M_z(E1)$ is positive and negative for $^{16}$O and $^{18}$O, respectively, indicating that protons distribute outward because of the Coulomb force. In this region, 
the dipole moments are comparable to those in $^{16}$O+$^{16}$O. 
In the $R \lesssim 9$ fm region, $M_z(E1)$ of $^{18}$O gradually increases with a decrease in $R$.
$M_z(E1)$ of $^{16}$O becomes positive and gradually increases. 
In this region, neutrons in $^{18}$O are attracted toward $^{16}$O because of nuclear interactions and push 
neutrons (or pull protons) in $^{16}$O. 
Although the finite dipole moments are observed in the calculations, the deviation of the distance between centers of mass of protons in $^{16}$O and $^{18}$O from the internuclear distance is quite small (less than 0.04 fm in the $R \geq 5$ fm region). Therefore, the dipole polarizations only have a minor effect on the $^{16}$O+$^{16,18}$O internuclear potentials.

The above discussions regarding the alignment and the dipole polarization of $^{18}$O is based on a strong-coupling scenario. 
In a weak-coupling scenario, these results suggest coupling with rotational members such as $J^\pi = 2^+$ state in the ground-state band 
contribute to the enhancement of sub-barrier fusion cross sections instead of coupling with $J^\pi = 1^-$ states, which is consistent with the CC model study\cite{PhysRevC.32.1432}.
In heavy and well-deformed systems, where the strong coupling picture works rather well in very low-energy reaction, deformation effects to near- and sub-barrier cross sections are discussed using the orientation-average of orientation-dependent cross section with sudden potentials\cite{PTP.128.1061}.
The alignment effects in the present results may be associated with the orientation-dependent cross sections. 

\section{Conclusions}
\label{conclusions}

We propose a method to obtain adiabatic internuclear potentials via energy variation with the intercluster-distance constraint in the AMD framework.
The potentials are applied to investigate the sub-barrier cross sections of $^{16}$O+$^{16}$O and $^{16}$O+$^{18}$O through a potential model.
For the MV1$'$ interaction, the theoretical cross sections agree with the experimental data, whereas for the D1S interaction, the theoretical cross sections are less than the experimental data.
Excitation of valence neutrons in $^{18}$O enhances sub-barrier fusion cross sections.
The alignment of deformed $^{18}$O is a dominant excitation effect, while dipole polarization effects are relatively weak.
 To understand sub-barrier fusion reactions, the details of the structural changes should be considered.
The present adiabatic internuclear potentials work well to describe sub-barrier nuclear fusions qualitatively.

\ack


 We thank Dr.~Hagino for providing  a code to calculate fusion cross sections with a potential model.
We also thank Prof.~Horiuchi, Prof.~Wada, Prof.~Tohsaki, and Dr.~Kimura for fruitful discussions.
Numerical calculations were conducted on the High-Performance Computing system at RCNP, Osaka University.
This study was supported by Grant-in-Aid for JSPS Fellows.
The study was also partly supported by Grant-in-Aid for Scientific Research from JSPS.

\bibliographystyle{ptephy}
\bibliography{potential_v3.2}

%
%

\end{document}